\documentclass[a4paper]{article}
\usepackage[pdftex]{graphicx}
\usepackage{INTERSPEECH2021}
\usepackage{cite}
\usepackage{bm}
\usepackage{amsmath}
\usepackage{amsfonts}
\usepackage{multirow}
\usepackage{afterpage}

\title{Enrollment-less training for personalized voice activity detection}
\name{Naoki Makishima, Mana Ihori, Tomohiro Tanaka,\\ Akihiko Takashima, Shota Orihashi, Ryo Masumura}

\address{NTT Media Intelligence Laboratories, NTT Corporation, Japan}

\begin{document}

\maketitle
\begin{abstract}
  We present a novel personalized voice activity detection (PVAD) learning method that does not require enrollment data during training.
  PVAD is a task to detect the speech segments of a specific target speaker at the frame level using enrollment speech of the target speaker.
  Since PVAD must learn speakers' speech variations to clarify the boundary between speakers, studies on PVAD used large-scale datasets that contain many utterances for each speaker.
  However, the datasets to train a PVAD model are often limited because substantial cost is needed to prepare such a dataset.
  In addition, we cannot utilize the datasets used to train the standard VAD because they often lack speaker labels.
  To solve these problems, our key idea is to use one utterance as both a kind of enrollment speech and an input to the PVAD during training, which enables PVAD training without enrollment speech.
  In our proposed method, called enrollment-less training, we augment one utterance so as to create variability between the input and the enrollment speech while keeping the speaker identity, which avoids the mismatch between training and inference.
  Our experimental results demonstrate the efficacy of the method.
\end{abstract}

\section{Introduction\label{sec:intro}}
Voice activity detection (VAD) is a technique to classify an acoustic segment into speech or non-speech, which is an important frontend step in a wide range of tasks such as speaker verification~\cite{TKinnunen2013_VADSV,YJung2019_VADSV}, emotion estimation~\cite{FRiengeval2014_VADEmotion}, and automatic speech recognition~\cite{TYoshimura2020_VADASR}.
Although many strategies have been proposed for VAD such as time-domain-energy-based methods and likelihood-ratio-based methods~\cite{LRebiner1975_bell,JSohn1999_statisticalVAD,ALee2004_GMMVAD,JRamirez2005_statisticalVAD}, fully neural network based methods have shown promising performance even under low signal-to-noise ratio (SNR) environments~\cite{XZhang2013_VAD,THughes2013_RNNVAD,FEyben2013_VADHollywood,NRyant2013_VADYoutube,SThomas2014_VAD,RMasumura2019_VAD,RLin2019_VAD,JLee2020_VAD}.

The VAD system addressed in these studies is trained so as to detect all speech segments regardless of the speaker, which is referred to as standard VAD.
On the other hand, studies have also looked at a VAD system that detects the voice activity of a specific target speaker at the frame level, which is referred to as personalized VAD (PVAD)~\cite{DGerber2017_multichannelTSVAD,SDing2018_personalVAD, IMedennikov2020_TSVAD}.
Compared with standard VAD, one strength of PVAD is that it does not detect the voice included in background noise that often leads to an unexpected response or an error of the downstream tasks in a real environment.
Moreover, since PVAD models can distinguish the speaker of the speech signals, they are suitable for downstream tasks that are intended to respond to a specific user, e.g., a speech recognition system that responds only to the user's voice.

In PVAD, an utterance-level target speaker embedding such as an i-vector~\cite{NDehak2010_ivector} or a d-vector~\cite{EVariani2014_dvector,LWan2018_dvector} is required to determine whether the input frame is the target speaker speech~\cite{SDing2018_personalVAD, IMedennikov2020_TSVAD}.
The target speaker speech to obtain such an embedding is called enrollment speech.
Note that a delay in the PVAD is almost the same as that of standard VAD because PVAD does not require the utterance-level information of the \textit{input speech} unlike the systems that directly combine standard VAD and the extraction of detected speech features~\cite{EVariani2014_dvector,LWan2018_dvector,DSnyder2018_Xvector} for speaker verification.

The enrollment speech is directly set by the user in inferences.
On the other hand, to use such speech during training, we have to prepare large-scale datasets that contain  many utterances for each speaker, and this is referred to as enrollment-full data.
Since speech characteristics such as the intonation or tone vary a little depending on the content of utterances, these utterances are used to learn such diversities, which is crucial to decide the boundaries between speakers.
However, preparing enrollment-full datasets in real environments requires substantial cost.
In addition, we cannot utilize real environment datasets used to train the standard VAD for training of the PVAD model because they often lack speaker labels and cannot provide enrollment speech, which is referred to as enrollment-less data.
Thus, datasets to train a PVAD model are often limited, and using such easily available enrollment-less datasets are indispensable for training of a robust PVAD model.

To solve these problems, we design a PVAD model that is trained without enrollment speech, i.e., we can train our PVAD model only with abundantly available enrollment-less datasets.
Our key idea is to use one utterance simultaneously for both an enrollment speech and an input to the PVAD during training.
Obviously, this strategy causes a mismatch between training and inference because the target speaker speech in the input and the enrollment speech are different when inferring in a real environment.
Therefore, overfitting must be avoided, and the diversity of the target speaker embeddings needs to be learned.

In this paper, we present a new PVAD training method that does not use enrollment speech during training, which is referred to as enrollment-less training.
To learn the diversity of the target speaker embeddings  from enrollment-less datasets, we introduce enrollment augmentation that combines SpecAugment~\cite{DPark2019_SpecAugment} and dropout~\cite{NStrivastava2014_dropout} to the target speaker embeddings.
The key point of the enrollment augmentation is that we should keep the speaker identity while producing the diverse target speaker embeddings.
We experimentally show that the combination of SpecAugment and dropout creates diverse target speaker embeddings resembling the ones generated from many utterances of the target speaker in terms of cosine similarity, which is often used to calculate the similarity between speaker embeddings~\cite{EVariani2014_dvector,ANagrani2020_SpeechEmbeddingCrossModal}.
To the best of our knowledge, this is the first study that trains PVAD without using enrollment speech.
Our experimental results show that the PVAD model trained with the proposed method using enrollment-less datasets achieves higher performance than the one trained with the conventional method using enrollment-full datasets.
Moreover, we experimentally show that the enrollment augmentation is crucial for the proposed enrollment-less training.

\section{Conventional methods}
\subsection{Standard voice activity detection}
We denote the time-frequency representation of an input and its standard VAD state sequence as $\bm X=(\bm x_1,\ldots, \bm x_T)$ and $\bm s=(s_1,\ldots, s_T)$, respectively, where $T$ denotes the length of the input, $\bm x_t\in\mathbb{R}^D$ denotes the $t$th frame of the input such as log mel-scale filterbank coefficients, $D$ denotes its dimension, and $s_t\in\mathbb{R}$ represents that the $t$th frame is the speech state when $s_t=1$ and that it is the non-speech state when $s_t=0$.
Standard VAD demands the estimation of the conditional probability of $\bm s$ given $\bm X$ in a frame-by-frame manner as
\begin{align}
P(\bm s|\bm X, \theta) = \prod_{t=1}^T P(s_t|\bm x_1,\ldots, \bm x_t,\theta)\label{eq:conditional_prob},
\end{align}
where
\begin{align}
P(s_t|\bm x_1,\ldots, \bm x_t,\theta) = [F(\bm x_1,\ldots,\bm x_t;\theta)]_{s_t},
\end{align}
$F(\cdot)\in\mathbb{R}^2$ denotes the neural network based function that outputs the conditional probability of each state, $\theta$ denotes its parameter, and $[\cdot]_{s_t}$ is the $s_t$th element of the vector.
A long short-term memory (LSTM) network is often used as $F(\cdot)$ to handle long-term dependencies between input features~\cite{STong2016_VADLSTM}.
Cross entropy (CE) is often used as a loss function to estimate the parameter $\theta$ that maximizes \eqref{eq:conditional_prob} as
\begin{align}
  \mathcal{L}_{\mathrm{CE}} = -\frac{1}{|\mathcal{D}|}\sum_{(\bm X, \bm s)\in\mathcal{D}}\sum_{t} \log P(s_t|\bm x_1,\ldots, \bm x_t,\theta),\label{eq:train_theta}
\end{align}
where $\mathcal{D}$ denotes the training minibatch for standard VAD training, and $|\mathcal{D}|$ denotes its size.
\subsection{Personalized voice activity detection}
We denote the time-frequency representation of an input, an enrollment speech, and the corresponding PVAD state sequence of the input as $\tilde{\bm X} = (\tilde{\bm x}_1, \ldots, \tilde{\bm x}_T)$, $\bm Y = (\bm y_1, \ldots, \bm y_J)$, and $\bm q=(q_1,\ldots, q_T)$, respectively, where $\tilde{\bm x}_t\in\mathbb{R}^D$ is the $t$th frame of the input acoustic feature for PVAD,
$\bm y_j\in\mathbb{R}^D$ is the $j$th frame of the acoustic feature of the enrollment speech, $J$ denotes the enrollment speech length, and $q_t\in\mathbb{R}$ represents that the $t$th frame is the target speech state when $q_t=1$ and that it is the non-speech state or the non-target speech state when $q_t=0$.
Note that $\tilde{\bm X}$ contains the speech of both the target speaker and the non-target speaker to learn to avoid detecting non-target speaker speech.
PVAD estimates the conditional probability of $\bm q$ given $\tilde{\bm X}$ and $\bm Y$ as
\begin{align}
P(\bm q|\tilde{\bm X}, \bm Y, \theta_{\mathrm{P}}) = \prod_{t=1}^T P(q_t|\tilde{\bm x}_1,\ldots, \tilde{\bm x}_t, \bm Y, \theta_{\mathrm{P}})\label{eq:conditional_prob_pvad},
\end{align}
where
\begin{multline}
P(q_t|\tilde{\bm x}_1,\ldots, \tilde{\bm x}_t, \bm Y, \theta_{\mathrm{P}}) =\\ [\mathrm{PVAD}(\tilde{\bm x}_1,\ldots,\tilde{\bm x}_t, \bm e_{\mathrm{target}};\theta_{\mathrm{P}})]_{q_t},\label{eq:pvad}
\end{multline}
\begin{align}
\bm e_{\mathrm{target}} = \mathrm{SpeakerModel}(\bm Y;\theta_{\mathrm{fix}}),\label{eq:speaker_model}
\end{align}
$\mathrm{PVAD}(\cdot)$ denotes the PVAD model, $\theta_{\mathrm{P}}$ denotes its parameter, $\bm e_{\mathrm{target}}\in\mathbb{R}^{K}$ denotes the target speaker embedding, $K$ denotes its dimension, $\mathrm{SpeakerModel}(\cdot)$ denotes the pretrained speaker model, and $\theta_{\mathrm{fix}}$ denotes its parameter that is pretrained for speaker verification.
Note that $\theta_{\mathrm{fix}}$ is often fixed during the training of PVAD.
CE is often used in PVAD as a loss function to estimate the parameter $\theta_{\mathrm{P}}$ that maximizes \eqref{eq:conditional_prob_pvad} as
\begin{align}
  \mathcal{L}_{\mathrm{CEP}} = -\frac{1}{|\mathcal{D}_{\mathrm{P}}|}\sum_{(\tilde{\bm X}, \bm Y,\bm q)\in\mathcal{D}_{\mathrm{P}}}\sum_{t} \log P(q_t|\tilde{\bm x}_1,\ldots, \tilde{\bm x}_t,\bm Y, \theta_{\mathrm{P}}),\label{eq:train_theta_pvad}
\end{align}
where $\mathcal{D}_{\mathrm{P}}$ denotes the training minibatch for the PVAD training.

\section{Proposed PVAD}
\subsection{Strategy\label{sec:strategy}}
In most personalized voice activity detection, a substantial amount of enrollment speech is required during training to learn the diversity of utterances and the boundary between speakers.
However, as described in section~\ref{sec:intro}, it limits the available dataset used for PVAD, which degenerates the PVAD performance.
Our method addresses this problem using one utterance simultaneously for both an enrollment speech and an input during training, i.e., we use the standard VAD dataset $\mathcal{D}$ instead of $\mathcal{D}_{\mathrm P}$ to train the PVAD model.

\begin{figure}
  \begin{center}
  \includegraphics[width=0.95\columnwidth]{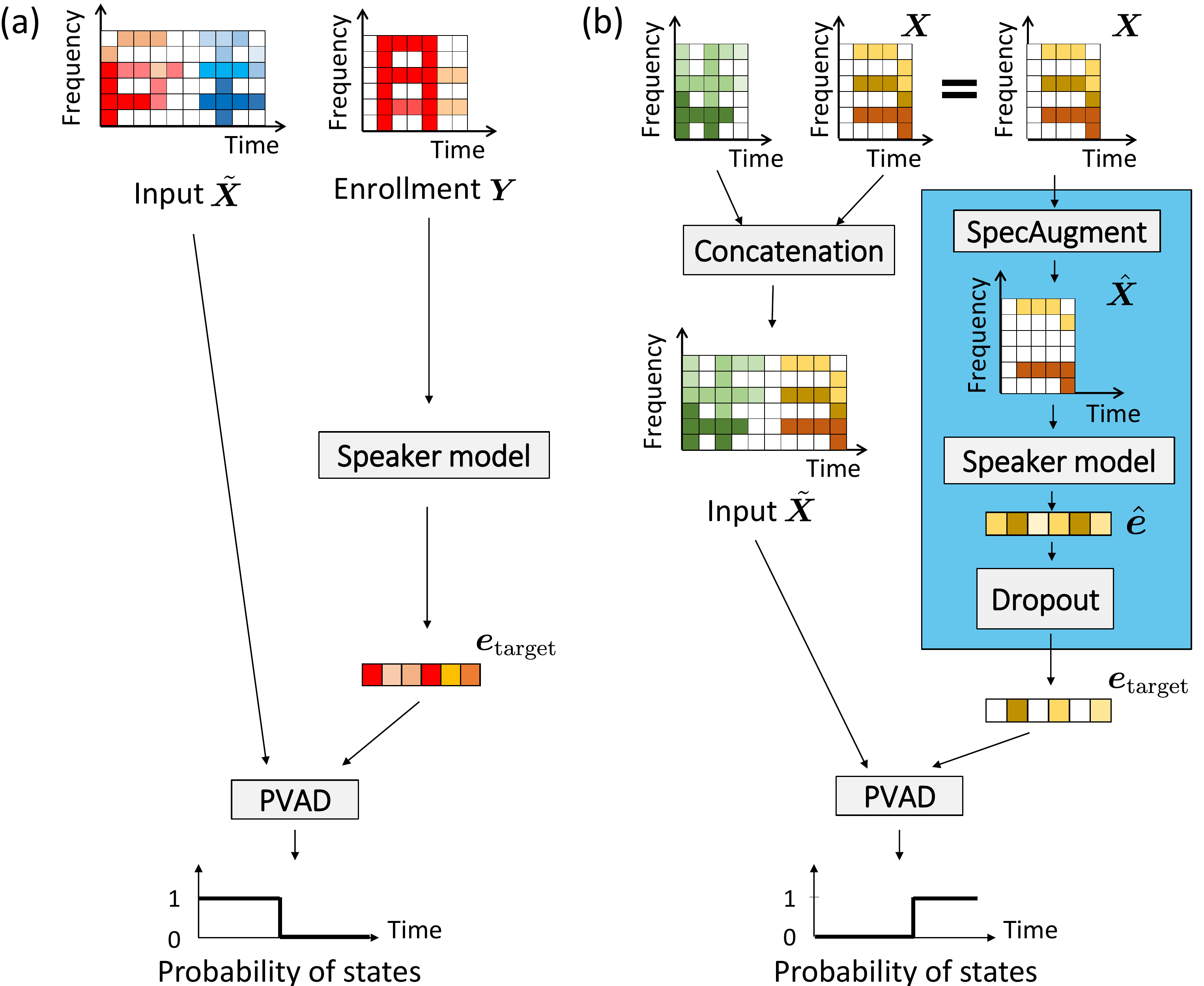}
  \vspace{-15pt}
  \end{center}
  \caption{Overview of proposed method in (a) inference and (b) training.
  Red, blue, green, and yellow slots in acoustic feature represent speech of different speakers.}
  \vspace{-18pt}
  \label{fig:overview}
\end{figure}

In inferences,  we use enrollment speech to estimate $\bm q$ by \eqref{eq:pvad} and \eqref{eq:speaker_model}, as shown in Fig.~\ref{fig:overview} (a).
On the other hand, since enrollment speech is not available during training, we simulate the PVAD training as follows.
First, multiple utterances in the standard VAD dataset are concatenated to create input $\tilde{\bm X}$, which includes multiple speakers' speech, as shown in  Fig.~\ref{fig:overview} (b).
Next, we consider one utterance before concatenation as target speaker speech, and we use it instead of enrollment speech, i.e., the target speech and the enrollment speech are identical.
Since this training causes the mismatch between training and inference, we augment the enrollment speech using enrollment augmentation, which is shown as a blue block in Fig.~\ref{fig:overview} (b).

The enrollment augmentation consists of three procedures: SpecAugment~\cite{DPark2019_SpecAugment}, the estimation of the target speaker embedding, and dropout~\cite{NStrivastava2014_dropout}.
First, we mask random frequency bands of the speech using SpecAugment, which enables the downstream speaker model to output the diverse speaker embeddings, each focusing on particular frequencies.
Then,  a pretrained speaker model outputs the speaker embedding of the masked speech.
Finally, dropout is applied to the speaker embedding received from the speaker model.
Since dropout is effective to prevent the overfitting~\cite{NStrivastava2014_dropout}, we use it to prevent PVAD from learning that the target speaker embedding is created from part of the input speech.
Although this enrollment augmentation is crucial for the proposed method, we assume that it is effective even when the enrollment-full dataset $\mathcal{D}_{\mathrm P}$ is used during training.
This is because the role of the augmentation is to create the speech variation and it can further increase the variation of enrollment speech in $\mathcal{D}_{\mathrm P}$.
We discuss it in section~\ref{sec:exp_sim}.

\subsection{Enrollment-less training\label{sec:train}}
In this section, we formulate the training of the proposed method.
Since the enrollment speech $\bm Y$ is not available during training, we use one utterance $\bm X$ in the standard VAD dataset to obtain $\bm e_{\mathrm{target}}$.
As described in section~\ref{sec:strategy}, we create $\tilde{\bm X}$ and $\bm q$ by concatenating multiple utterances and their state sequences w.r.t. the target speaker in the enrollment-less dataset.
In our method, PVAD is trained to estimate the conditional probability of $\bm q$ given $\tilde{\bm X}$ and $\bm X$ as
\begin{align}
P(\bm q|\tilde{\bm X}, \bm X,\theta_{\mathrm P}) = \prod_{t=1}^T P(q_t|\tilde{\bm x}_1,\ldots, \tilde{\bm x}_t, \bm X,\theta_{\mathrm P}).\label{eq:conditional_prob_prop}
\end{align}

First, SpecAugment~\cite{DPark2019_SpecAugment} is applied to $\bm X$ to obtain the frequency masked speech $\hat{\bm X}$.
Given $\hat{\bm X}$, the speaker model outputs the speaker embedding as
\begin{align}
  \hat{\bm e} = \mathrm{SpeakerModel}(\hat{\bm X};\theta_{\mathrm{fix}}).
\end{align}
We obtain $\bm e_{\mathrm{target}}$ by applying the dropout of ratio $p$ to $\hat{\bm e}$.
Given $\bm e_{\mathrm{target}}$, the conditional probability of $\bm q$ is obtained as
\begin{multline}
  P(q_t|\tilde{\bm x}_1,\ldots, \tilde{\bm x}_t, \bm X, \theta_{\mathrm{P}}) = \\
[\mathrm{PVAD}(\tilde{\bm x}_1,\ldots,\tilde{\bm x}_t, \bm e_{\mathrm{target}};\theta_{\mathrm{P}})]_{q_t}.
\end{multline}

We denote the dataset created from $\mathcal{D}$ as $\mathcal{D}'$.
The loss function to train $\theta_{\mathrm P}$ is obtained as
\begin{align}
\mathcal{L} &= -\frac{1}{|\mathcal{D}'|}\sum_{(\tilde{\bm X}, \bm X, \bm q)\in\mathcal{D}'}\sum_t \log P(q_t|\tilde{\bm x}_1,\ldots,\tilde{\bm x}_t, \bm X,\theta_{\mathrm P}).\label{eq:weighted_loss}
\end{align}
\subsection{Implementation}
This section describes the architecture of the pretrained speaker model and the PVAD model.
The speaker model consists of three bidirectional LSTM (BLSTM) layers with 256 units.
The forward and backward output of the BLSTM are concatenated.
The third BLSTM output is averaged using an attention mechanism that calculates the weighted mean of the frame-level features~\cite{KOkabe2018_Attentivepooling} to get the target speaker embedding.

The input to the PVAD model is a concatenation of the target speaker embedding and the input acoustic feature in a frame-wise manner.
The PVAD model consists of four LSTM layers followed by a dropout with a ratio of 0.5, one linear layer, and softmax activation.
The number of units of all the LSTM layers and the linear layer is 256 and 2, respectively.
\begin{table*}[t!]
  \begin{minipage}[t!]{\textwidth}
  \caption{Evaluation results using clean test data.}
    \vspace{-20pt}
  \begin{center}
  \begin{tabular}{|l|ccc|} \hline
    \textbf{Method} & \textbf{AP (ns/nts)} & \textbf{AP (ts)} & \textbf{mAP} \\ \hline
    VAD              & (0.926)     & (0.543)     & (0.691)          \\
    Conventional PVAD (enroll-full w/o aug.)  & 0.965 & 0.769 & 0.904 \\
    PVAD (enroll-full w/ aug.)   & 0.970 & 0.814 & 0.919 \\
    PVAD (enroll-less w/o aug.) & 0.943 & 0.779 & 0.899   \\
    Proposed PVAD (enroll-less w/ aug.)  & \textbf{0.990} & \textbf{0.937} & \textbf{0.970} \\
    \hline
  \end{tabular}
  \label{table:res_clean}
  \end{center}
\end{minipage}
\bigskip
\begin{minipage}[t!]{\textwidth}
  \caption{Evaluation results using noisy test data.}
  \vspace{-20pt}
  \begin{center}
    \begin{tabular}{|l|ccc|ccc|}\hline
      \multirow{2}{*}{\textbf{Noise}}& \multicolumn{3}{|c|}{Conventional PVAD (enroll-full w/o aug.)} & \multicolumn{3}{|c|}{Proposed PVAD (enroll-less w/ aug.)} \\ \cline{2-7}
                  &\textbf{AP (ns/nts)} & \textbf{AP (ts)} & \textbf{mAP}&\textbf{AP (ns/nts)} & \textbf{AP (ts)} & \textbf{mAP}\\\hline
              5 dB   &  0.932 & 0.696 &0.840&\textbf{0.970}&\textbf{0.856}&\textbf{0.931} \\
              10 dB  & 0.947&0.735&0.867&\textbf{0.979}&\textbf{0.894}&\textbf{0.946}\\
              15 dB  & 0.956&0.756&0.886&\textbf{0.984}&\textbf{0.916}&\textbf{0.955}\\
              20 dB  & 0.974&0.771&0.895&\textbf{0.986}&\textbf{0.928}&\textbf{0.960}\\ \hline
  \end{tabular}
  \label{table:res_noisy}
  \end{center}
  \end{minipage}
  \vspace{-10pt}
\end{table*}
\section{Experiments\label{sec:exp}}

\subsection{Dataset}
We used a large-scale Japanese speech dataset for an evaluation.
We prepared a home-made 380-hour enrollment-less speech dataset $\mathcal{D}$ and a home-made 90-hour enrollment-full  speech dataset $\mathcal{D}_{\mathrm P}$ for the training data.
The validation data of the enrollment-less dataset and enrollment-full dataset consisted of 25 hours and 5 hours of data, respectively.
The enrollment-full dataset was used to train the PVAD model with the conventional method~\cite{SDing2018_personalVAD}.
Although the enrollment-full dataset can also be used for the proposed method, we used only the enrollment-less dataset to train the model with the proposed method for a simple comparison.
The number of speakers in enrollment-full data was 256 and the number of utterances per speaker was 300 on average.
We created a noisy speech dataset by synthesizing 120 manually-constructed noise types (car, shopping mall, factory, etc.) into the clean speech data.
SNR levels were randomly varied between 5 to 30 dB.
We used both clean speech and noisy speech data for the training.

We prepared five thousands pairs of clean input, the enrollment speech, and the PVAD state sequences for the testing.
The average length of one utterance of a speaker was 4 seconds, and the input included one to three speaker utterances without overlapping.
The length of the enrollment speech was 8 seconds on average, and the sentences were not included in the input.
We also prepared a noisy speech dataset by corrupting the clean input with two unseen noise types (a crowd and a train station) at four noise levels: i.e., with SNR values of 5, 10, 15, and 20 dB.
The sampling rate of the entire dataset was 16 kHz.

\subsection{Settings}
First, we compared the proposed PVAD trained using the enrollment-less dataset (enroll-less w/ aug.) and the conventional PVAD trained using the enrollment-full dataset (enroll-full, w/o aug.), where aug. means the enrollment augmentation in section~\ref{sec:train}.
Next, we also evaluated the PVAD model trained using the enrollment-less dataset without enrollment augmentation (enroll-less w/o aug.) and the PVAD model using the enrollment-full dataset with enrollment augmentation (enroll-full w/ aug.) to investigate the effect of the enrollment augmentation.
PVAD (enroll-less w/o aug.) was trained using the target speaker embedding of $\bm X$ without SpecAugment or dropout, and PVAD (enroll-full w/ aug.) was trained by applying SpecAugment and dropout to enrollment speech $\bm Y$.
The standard VAD trained using the enrollment-less dataset was also evaluated just for reference.

We concatenated 1 to 3 randomly selected utterances when using the enrollment-less dataset for PVAD training.
We used 40 dimensional log mel-scale filterbank coefficients appended with delta and acceleration coefficients as acoustic features; they were extracted using a 20-ms-long Hamming window with a 10-ms-long shift.
The input to the speaker model and PVAD was the 840 dimensional acoustic features formed by concatenating the $\pm 3$ left-right contexts of the 120 dimensional acoustic features.
\begin{figure}[t]
  \begin{center}
  \includegraphics[width=1.0\columnwidth]{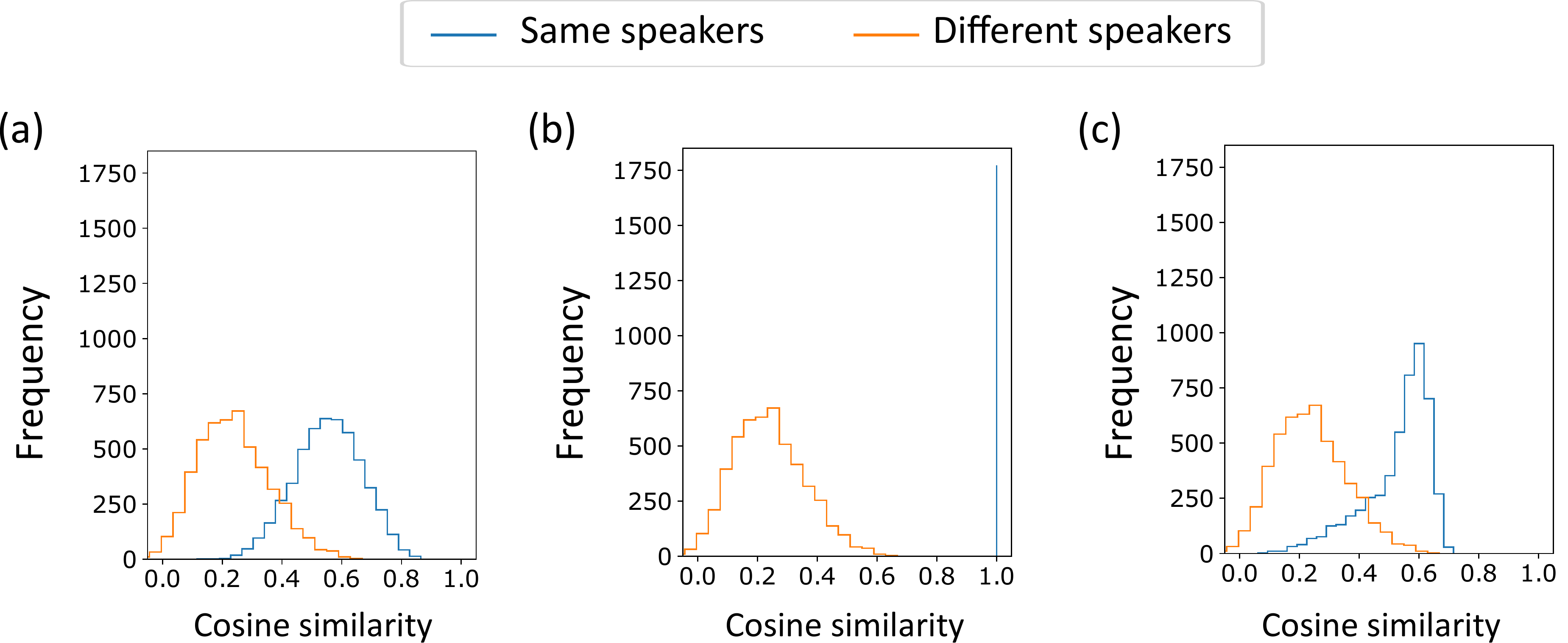}
  \vspace{-20pt}
  \end{center}
  \caption{(a) Cosine similarities of embeddings created using many enrollment utterances, (b) cosine similarities of embeddings without enrollment speech or enrollment augmentation, and (c) cosine similarities of embeddings created with enrollment augmentation without enrollment speech.
  }
  \vspace{-5pt}
  \label{fig:similarity}
\end{figure}
The speaker model was pretrained with a speaker classification of 5994 speakers using the VoxCeleb2 dataset~\cite{JChung2018_Voxceleb2}.
It was optimized using the Adam algorithm~\cite{DKingma2015_Adam} with a minibatch size of 128.
We set the initial learning rate of the algorithm to 0.001.
The training steps were stopped if the loss on the validation set did not decrease for firve epochs in succession.
We fixed the parameter of the speaker model during the training of the PVAD model.
The PVAD model was optimized in the same way as that of the speaker model except that we set the initial learning rate of the algorithm to 0.0001.
We randomly masked one-third of the acoustic features in SpecAugment.
We set the dropout ratio $p$ to $0.5$.

We used the average precision (AP) for each class and mean AP (mAP) over all the classes to evaluate performance.
We calculated the mAP using the micro-mean over all the classes as in \cite{SDing2018_personalVAD}.

\subsection{Diversity of target speaker embedding obtained with enrollment augmentation\label{sec:exp_sim}}
To examine whether enrollment augmentation is effective to create diverse target speaker embeddings while preserving the speaker characteristics, we compared the cosine similarities of speaker embeddings created using many enrollment utterances of a speaker and those created with enrollment augmentation from one utterance.
We also calculated the cosine similarities between different speaker embeddings without enrollment augmentation just as reference.
We used the clean test dataset.

Fig.~\ref{fig:similarity} (a) shows the  cosine similarities of embeddings created using many enrollment utterances, (b) shows the cosine similarities of embeddings without enrollment speech or enrollment augmentation, and (c) shows the cosine similarities of embeddings created with enrollment augmentation without enrollment speech.
We can see that the cosine similarities between the speaker embeddings of the same speakers formed a distribution around 0.5 and those of different speakers formed a distribution around 0.3, as shown in Fig.~\ref{fig:similarity} (a).
These distributions were formed because speech characteristics varied depending on the content of the utterance, and speaker embeddings were affected by the variance~\cite{SWang2017_whatSpeakerEmbeddingEncode,YLiu2018_SpeakerEmbeddingPhonetic,SWang2019_UsagePhoneticInformation}.
Fig.~\ref{fig:similarity} (b) shows that when no enrollment speech was available and when enrollment augmentation was not applied, speaker embedding of the same speaker did not vary, and a peak appeared at cosine similarity of 1.
By applying enrollment augmentation, we can see that the variations in the same speaker embeddings were created around cosine similarity of 0.5 without enrollment speech, as shown in  Fig.~\ref{fig:similarity} (c).
These results demonstrated that enrollment augmentation enabled us to create diverse speaker embeddings of a speaker even when we do not have the enrollment utterances.

\subsection{Detection performance}
First, we compared the performance using clean test data.
Table~\ref{table:res_clean} shows the results.
AP (ns/nts) represents the AP for non-speech or non-target speech, and AP (ts) represents the AP for target speech.
We show the performance of the standard VAD just as reference and put it in parentheses.
In comparing the results of the proposed PVAD (enroll-less w/ aug.) with the others, we can see that our proposed training method achieved the best performance.
The results show that although using data without speaker labels could degrade the performance of PVAD when we did not use enrollment augmentation (conventional PVAD (enroll-full w/o aug.) vs. PVAD (enroll-less w/o aug.)), the data became indispensable to improve the PVAD performance when we used enrollment augmentation.
We can also find that enrollment augmentation was effective even when enrollment-full dataset was used (conventional PVAD (enroll-full w/o aug.) vs. PVAD (enroll-full w/ aug.)).
This is probably because enrollment augmentation further increased the variation of enrollment speech in the enrollment-full dataset.

Next, we compared the performance using noisy test data.
Table~\ref{table:res_noisy} shows the result of the conventional method and the proposed one.
We averaged the results of two types of noises.
We can see that the proposed method detected the target speaker speech and avoided  detecting the non-target speaker speech or noise under noisy environments.

\section{Conclusions}
In this paper, we proposed a new PVAD learning method that does not need enrollment speech during training.
We achieved it using one utterance for both an input and an enrollment utterance.
Since the PVAD model must learn speakers' speech variations to clarify the boundary between speakers, we combined SpecAugment and dropout to create diverse target speaker embeddings while preserving the speaker identity from one utterance.
Our experimental results show that the PVAD model trained with the proposed method using enrollment-less datasets achieves higher performance than the one trained with the conventional method using enrollment-full datasets.

\bibliographystyle{IEEEtran}

\bibliography{reference}
\end{document}